\documentclass[dvips]{article}
\usepackage{icrctc07}

\title{Maximum Likelihood method for ultrahigh energy cosmic ray cross correlations with astrophysical sources}
\shorttitle{Maximum Likelihood method for UHECR correlations}
\authors{R. Jansson, G. R. Farrar.}
\shortauthors{R. Jansson and G. R. Farrar}
\afiliations{Center for Cosmology and Particle Physics,
 and Department of Physics\\ New York University, NY, NY 10003,USA}
\email{rj486@nyu.edu}

\abstract{We extend the Maximum Likelihood method used by HiRes to study cross correlations between a catalog of candidate astrophysical sources and Ultrahigh Energy Cosmic Rays (UHECRs), to allow for differing source luminosities. Our approach permits individual sources to be ranked according to their likelihood of having emitted the correlated UHECRs. We test both old and new method by simulations for various scenarios. We conclude that there are 9 true correlation between HiRes UHECRs and known BLLacs, with a $6\times 10^{-5}$ probability of such a correlation arising by chance.}

\begin{document}
\maketitle

\section{Introduction}
 
Hints of excess correlations have been reported between ultrahigh energy cosmic rays (UHECRs) and BLLacs \cite{TT:BLLacs01,TT:BLLacs04,HR:BLLacs} and between UHECRs and x-ray clusters \cite{pf05}. The analyses in refs. \cite{TT:BLLacs01,TT:BLLacs04,pf05} present the number of correlated events as a function of angular separation between UHECR and source, and give the associated ``chance probability'' of finding a correlation at the observed level as a function of angular separation, in a large number of simulations with no correlations.  

In order to incorporate the experimental resolution on an event-by-event basis, ref. \cite{HR:BLLacs} proposed a Maximum Likelihood-type procedure and applied it to studying correlations between UHECRs and BLLacs. This procedure (denoted the HiRes procedure, below) is motivated under the unphysical assumption that every candidate BLLac source has the same apparent luminosity. Even if BLLacs were standard candles with respect to UHECR emission, the BLLacs in the catalog have a large range of distances which would imply an even larger range of apparent luminosities, so one does not want to rely on such an assumption.  

In the present paper we introduce a ML prescription which avoids the assumption of equal apparent source luminosity and allows the potential sources to be ranked according to the probability that they have emitted the correlated UHECRs. We test and compare both methods with simulations. We find that the HiRes method gives the correct total number of correlated events even when the sources do not have equal apparent luminosities, as long as the numbers of events are sufficiently low and the candidate sources are not too dense or clustered themselves. In general, our new method performs better in these more challenging cases, but very occasionally can be ``tricked`` by some special configuration. We apply the new procedure to BLLacs.

\section{Maximum Likelihood methods for the cross-correlation problem}\label{method}

In the HiRes Maximum Likelihood method \cite{HR:BLLacs} the aim is to find, among $N$ cosmic ray events, the correct number of events, $n$, that are \emph{truly} correlated with some sources, of which the total number is $M$. Hence, there will be $N-n$ background events whose arrival directions are given by a probability density $R(\bf{x})$, which is simply the detector exposure to the sky as a function of angular position, $\bf{x}$. For a true event with arrival direction $\bf{s}$, the \emph{observed} arrival direction is displaced from $\bf{s}$ according to a probability distribution $Q_i(\bf{x},\bf{s})$.  For the analysis given in \cite{HR:BLLacs}, $Q_i$ is taken to be a $2d$ symmetric Gaussian of width equal to the resolution $\sigma_i$, of the $i$th event. (Note that, throughout, the parameter $\sigma$ we quote appearing in a $2d$ Gaussian is related to $\sigma_{68}$, the radius containing 68\% of the cases by $\sigma_{68} = 1.51\sigma$.)

The probability density of observing the $i$th  event in direction $\textbf{x}_i$ is 
$$ P_i(\textbf{x}_i) = \frac{n\sum_j^MQ(\textbf{x}_i,\textbf{s}_j)R(\textbf{s}_j)}{N\sum_{k=1}^{M}R(\textbf{s}_k)} + \frac{N-n}{N}R(\textbf{x}_i), $$
and the \emph{likelihood} for a set of $N$ events is defined to be $\mathcal{L}(n) = \prod_{i=1}^NP_i(\textbf{x}_i)$,
which is maximized when $n$ is the true number of correlated events. Since $\mathcal{L}$ is a very small number, which depends on the number of events, it is more useful to divide $\mathcal{L}$ by the likelihood of the \emph{null hypothesis}, i.e., $n=0$, to form the likelihood ratio $\mathcal{R}(n) = \mathcal{L}(n)/\mathcal{L}(0)$. The logarithm of this ratio is then maximized to obtain the number of correlated events, $n$.

To generalize this method to allow for sources with differing luminosities we assign a number of correlations, $n_j$, separately for each source, with $n = \sum_{j=1}^Mn_j$. The probability density generalizes to
\begin{eqnarray}
P_i(\textbf{x}_i) = \frac{\sum_j^Mn_jQ(\textbf{x}_i,\textbf{s}_j)R(\textbf{s}_j)}{N\bar{R}_s} \nonumber \\
+  \frac{N-\sum_j^Mn_j}{N}R(\textbf{x}_i),
\end{eqnarray}
where $\bar{R}_s=\sum_j^MR(\textbf{s}_j)/M$. Maximizing $\ln \mathcal{R}$ gives the set $\{n_j\}$ of $M$ numbers, containing the individual apparent source luminosities. We also get the set $\{(\ln \mathcal{R})_i\}$ of $N$ numbers, providing information about how strongly correlated the individual cosmic ray events are to the catalog of sources. 

A crucial difference of the generalized method compared to the HiRes method is that $n_{tot} = \sum n_j$ gives an estimate of \emph{all} correlations, i.e., both \emph{true} and \emph{random} correlations, whereas the HiRes method yields only an estimate of the number of \emph{true} correlations.   A crude estimate of the number of true correlations for the new method is $n_{tot}-\bar{n}_{rand}$, where $\bar{n}_{rand}$ is the average number of correlations obtained when cosmic rays are uncorrelated to the data set of potential sources. 

A better measure of $n_{true}$ is summarized by the following equations, where the superscript in parentheses labels the ``order'' of refinement, giving successively better approximations to $n_{true}$:
\begin{eqnarray}
n^{(0)} &=& n_{tot} = \sum_j^Mn_j\\
n^{(1)} &=& \bar{f}^{-1}(n^{(0)} - \bar{n}_{rand}(N))\\
n^{(2)} &=& \bar{f}^{-1}(n^{(0)} - \bar{n}_{rand}(N-n^{(1)})),
\end{eqnarray}
where $\bar{n}_{rand}(N)$ is the average total number of correlations found with $N$ randomly generated events, and where $\bar{f}$ is the average fraction of true events that are recovered as correlated. This fraction will typically be slightly smaller than unity since under the assumption that true correlations are separated according to a Gaussian distribution, some events are separated too far from their sources to be accepted as correlated events. We measure $\bar{f}$ by simulations.
\begin{figure}
 \begin{center}
 \includegraphics[scale=.6]{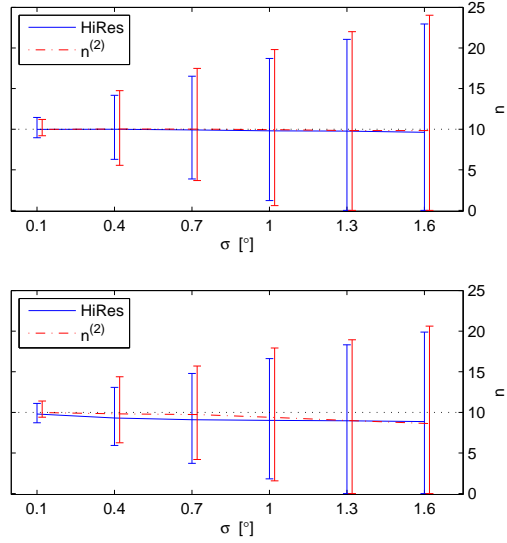}
 \end{center}
 \caption{The total number of correlations vs. the detector resolution. Error bars (slighly separated horizontally for readability) contain 90\% of the cases. \emph{Top:} Random sources; \emph{Bottom:} actual BLLac positions $-30^\circ \leq b \leq 90^\circ $.}
 \label{rand_bllacs}
 \end{figure}

\section{Testing with simulations}\label{simulations}

In vetting the two methods with simulations, we test their ability to correctly reproduce the number of true correlations on mock data sets. This allows us to explore the effect of large event and source densities, the effect of anisotropy in the source distribution, and the consequences of having incorrect event resolutions.

\begin{figure}
\includegraphics[scale=.5]{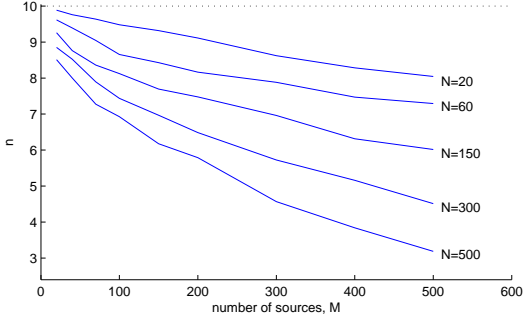}
\caption{Correlations found by HiRes method with 10 true correlations, $M$ sources and $N$ events in 100 square degrees.}
\label{dense} 
\end{figure}

We begin by testing the ability to reproduce the number of true correlations in the simplest case of dilute, random sources. A first simulation is done using half the sky, uniform detector exposure $R(\textbf{x})$, 156 randomly distributed sources and 271 cosmic rays (the numbers relevant to the BLLac studies of \cite{TT:BLLacs04} and \cite{HR:BLLacs}). Ten of the cosmic rays are Gaussianly aligned (a cosmic ray paired to a source, with angular separation according to the probability density $Q$, taken to be a $2d$ Gaussian of width $\sigma$), for various event resolutions. A second simulation uses the actual BLLacs source positions; these are more clustered than the random case. As shown in figure \ref{rand_bllacs}, both methods reproduce well on average the correct number of correlations, with similar error bars (which include 90\% of the 10k realizations). As $\sigma$ increases, the dispersion in the number of found correlations increases rapidly.

For samples with very large numbers of events and of potential sources, we expect the generalized ML method to perform worse than the HiRes method, beacause $n^{(2)}$ is obtained by taking the difference of two very large numbers, $n^{(0)}$ and $\bar{n}_{rand}$. Moreover, as the source density becomes very large it becomes impossible to reliably distinguish the ``contributing'' sources. The total number of true correlations becomes the only interesting quantity to calculate. Thus, only the HiRes method should be used for the case of very high event densities. However, the HiRes method also deteriorates at high densities, as shown in figure \ref{dense}.

\begin{figure}
 \begin{center}
 \includegraphics[scale=.56]{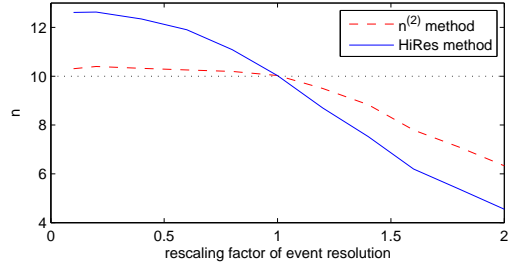}
 \end{center}
 \caption{Average number of found correlations as a function of the factor by which $\sigma$ is rescaled by when ``Gaussianly aligning'' sources.}
 \label{skew}
 \end{figure}

If the resolution of cosmic ray events are consistently over- or underestimated in a given data set, the extracted correlations will be incorrect. To test the sensitivity of the two methods to this problem we repeat the first type of simulations, but rescale the event resolution when aligning a cosmic ray to a source. In figure \ref{skew} the average number of found correlations for the two methods are plotted as a function of the amount by which $\sigma$ is rescaled. The new method is far less sensitive to incorrectly estimated resolution than is the HiRes method.

\begin{figure}
\includegraphics[scale=.7]{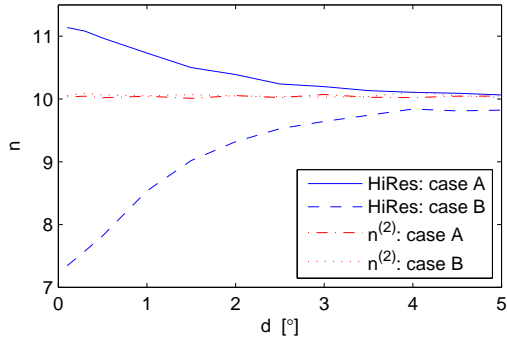}
\caption{Sensitivity to clustering in source dataset, for UHECRs from (A) dense or (B) sparse regions.}
\label{clustering} 
\end{figure}

Significant spatial correlations within the data set of potential sources may skew the found number of UHECR correlations. In figure \ref{clustering} we show the results of a simulation with clustering of potential sources introduced by hand. The figure shows the mean for 10k realization of 271 cosmic ray events with $\sigma = 0.4$, and two different scenarios for the correlation with source clustering, for 156 candidate sources.   In both cases, candidate sources and CRs are distributed over one hemisphere; ten clusters of candidate sources are randomly distributed on the sky, each consisting of ten individual candidate sources distributed around the cluster center according to a $2d$ Gaussian of width $d$ degrees. The remaining 56 candidate sources are placed at random in the hemisphere. In the first case, a randomly picked candidate source in each cluster has one cosmic ray event Gaussianly aligned to it and the remaining 261 cosmic rays are placed at random. In the second case, ten CRs are Gaussianly aligned with ten of the randomly  placed source candidates and the remaining CRs are placed at random. As figure \ref{clustering} demonstrates, the HiRes method significantly overestimates the true number of correlations if the sources are in clustered regions and underestimates it when the candidate sources show significant clustering but the UHECRs do not come from the clustered regions. By contrast the new method performs well in this test.

\section{Application to BLLacs}\label{applications}

The binned analysis performed in \cite{TT:BLLacs04} on the sample of 156 BLLacs with optical magnitude $m<18$ from the Veron 10th Catalog \cite{veron10} and the 271 HiRes events with $E>10^{19}$ eV showed a correlation at the $10^{-3}$ level. This was subsequently corroborated using the HiRes method \cite{HR:BLLacs}, with $n=8.0$ found correlations; the fraction of Monte Carlo runs with greater likelihood than the real data, was found to be $2\times 10^{-4}$. Using the generalized ML method we find the number of correlations to be $n^{(2)}=9.2$, with $\mathcal{F}=6\times 10^{-5}$. As shown in figure \ref{rand_bllacs}, the HiRes method underestimates the correct number of correlations by 0.7 on average in these conditions. However, the difference in these results is consistent with the dispersion in simulations. For a list of the BLLacs correlated to UHECRs and further analysis, see forthcoming paper.

\section{Conclusions}\label{summary}

We have introduced a generalization of the HiRes Maximum Likelihood method, which allows the most likely sources of individual events to be identified and ranked.  Using simulations we have tested the two Maximum Likelihood methods and find that they complement each other well: the HiRes method allows a fast way to estimate the number of true correlated events, while the new method gives the quality of correlation between individual sources and cosmic rays rather than just the total number of correlated events.  Furthermore, the new method is less sensitive to the validity of the estimated angular resolution, and is better when candidate sources are clustered (as BLLacs are). We conclude that both methods should be used; if they disagree markedly on the total number of correlations the data sets may be aberrant.  Applying the new method to the case of BLLacs, confirms an excess of correlations between HiRes cosmic rays with $E>10^{19}$ eV \cite{HR:BLLacs} and BLLacs of the Veron 10th catalog.  The excess obtained is slightly larger (9.2 rather than 8.0 events) than with the HiRes method, but the two values are compatible within the range of variation found in simulations.

\section{Acknowledgements}
We thank S. Westerhoff and C. Finley for information and discussions.  This research has been supported in part by NSF-PHY-0401232.

\bibliography{icrc1234}
\bibliographystyle{plain}

\end{document}